# Determination of the elastic moduli of CVD graphene by probing graphene/polymer Bragg stacks

**Bohai Liu[1, 2], Christos Pavlou[3, 4], Zuyuan Wang[2], Yu Cang[5], Costas Galiotis[3, 4, *] and George Fytas[2, *]**

[1] Center for Phononics and Thermal Energy Science, China-EU Joint Center for Nanophononics, School of Physics Science and Engineering, Tongji University, 200092 Shanghai, PR China
[2] Max Planck Institute for Polymer Research, Ackermannweg 10, 55128 Mainz, Germany
[3] Institute of Chemical Engineering Sciences, Foundation of Research and Technology-Hellas (FORTH/ICE-HT), Stadiou Street, 26504 Platani, Patras, Greece
[4] Department of Chemical Engineering, University of Patras, 26504 Patras, Greece
[5] School of Aerospace Engineering and Applied Mechanics, Tongji University, 100 Zhangwu Road, 200092 Shanghai, PR China
**Email:** c.galiotis@iceht.forth.gr, galiotis@chemeng.upatras.gr (C. Galiotis) and fytas@mpip-mainz.mpg.de (G. Fytas)

## Abstract

Graphene has been widely used in the form of micro-flakes to fabricate composite materials with enhanced mechanical properties. Due to the small size of the inclusions and their random orientation within the matrix, the superior mechanical properties of graphene cannot be fully exploited. Recently, attempts have been made to fabricate nanolaminate composites by interleaving large sheets of chemical vapor deposition (CVD) monolayer graphene between thin layers of polymer matrices. However, CVD graphene is inevitably accompanied by wrinkles that are formed in the synthesis process, and it remains unknown how the wrinkles affect the mechanical properties of graphene. Here, we employ Brillouin Light Spectroscopy (BLS) to study the elastic moduli of CVD graphene by probing graphene/poly(methylmethacrylate) hybrid Bragg stacks at zero strain. We find the Young's and shear moduli of the CVD graphene, which has wrinkles in the form of sharp elevations of height of about 6 nm and a FWHM of ca. 30 nm, to be 680 ± 16 and 290 ± 10 GPa, respectively, with the former being about 30% lower than that of exfoliated, flat graphene. This work sheds light on the elastic properties of CVD graphene and provides a method that can be extended to studying the wrinkle-induced softening effect in other two-dimensional materials.

# 1. Introduction

The unique hexagonal sp$^2$ C-C bonds endow graphene with remarkable physical properties, including extremely high thermal conductivity [1], electron mobility [2], and intrinsic strength [3]. Ever since its first fabrication by mechanical exfoliation [4], graphene has attracted extensive attention in the materials research community. For the mechanical properties, exfoliated suspended graphene has a Young's modulus of 1 ± 0.1 TPa, and a breaking strength of 42 N/m [3]. On the theoretical side, the Young's modulus of 1050 GPa and tensile strength of 110 GPa were reported [5].

The superior mechanical properties of graphene have made it an ideal filler to reinforce polymer materials [6-9]. To realize reliable functional materials, mass production and large-scale integration of graphene are prerequisites. Fortunately, chemical vapor deposition (CVD) has emerged as a promising solution [10-12], which perfectly satisfies the requirements and has been widely applied in graphene-based composites [13, 14]. However, multiple grain boundaries and atomic defects in CVD graphene can deteriorate its properties [15], such as the intrinsic strength, as the motion of dislocations is interrupted by these defects [16]. Furthermore, due to the thermal expansion mismatch between graphene and the copper substrate, severe biaxial stresses are developed on CVD graphene as it cooled from approximately 1000 $^o$C down to ambient temperature. These stresses are relaxed by the formation of a network of folds that form a mosaic structure as seen in many AFM images [17, 18]. Additionally, CVD graphene typically replicates the topography of the substrate. All of these can inevitably introduce out-of-plane wrinkles [19-22].

The elastic properties of CVD graphene have been theoretically and experimentally investigated. Theoretically, it has been shown that wrinkles play a dominant role in softening CVD graphene membranes [23]. The wrinkle-induced effect has been experimentally measured by Raman spectroscopy [24] and nanoindentation using atomic force microscopy (AFM) [25]. The latter addresses a small area under the AFM tip, and the highly non-uniform strain distribution over the sample can result in large spatial fluctuations of the two-dimensional elastic modulus. The former determines the strain-induced shift of the characteristic G and 2D Raman peaks of graphene [26, 27], and this allows the estimation of the elastic modulus via the Gruneisen parameters and the Poisson's ratio of the substrate [28]. For CVD graphene, however, the Raman wavenumber shifts with strain for both G and 2D peaks are markedly lower than those obtained from exfoliated graphene which indicate that the CVD modulus must be lower than the quoted modulus of ~1 TPa for monolayer graphene [27].

Despite the several aforementioned studies on the elastic properties of CVD graphene, a comprehensive understanding of the wrinkle-induced softening effect on the elasticity is still lacking. To tackle this problem, noncontact measurements of large-area CVD graphene appear to be an effective way for a quantitative study, particularly at zero strain. To the best of our knowledge, no such study has been reported in the literature. As a non-contact, non-destructive technique, Brillouin light spectroscopy (BLS) detects thermally excited phonons in the gigahertz (GHz) range, whose frequencies are directly related to the elastic moduli. It has been utilized to determine the elastic modulus of laminated materials, including poly(methyl methacrylate) (PMMA)/$SiO_2$ Bragg stacks [29], clay/polymer Bragg stacks [30], and PC/PMMA multilayer films [31]. Here, we employ BLS to study the wrinkle-introduced effect on the elastic moduli of CVD graphene by probing CVD graphene/PMMA (Gr/PMMA) Bragg stacks. From the BLS measurements, we directly obtained the sound velocities of the longitudinal acoustic phonons for both PMMA and the Gr/PMMA stacks. We conducted phononic band structure calculations using the Young's modulus of the CVD graphene as an adjustable parameter. We also performed tensile tests to compute the stiffness of the Gr/PMMA stacks in the axial direction at moderate strains and compared the results with the modulus from the BLS experiments.

## 2. Methods

### 2.1 Preparation and characterization of CVD Gr/PMMA multilayer stacks

Graphene growth was performed in a commercial CVD reactor (AIXTRON Black Magic Pro, Germany) on 7 cm × 7 cm copper sheets (JX Nippon Mining & Metals, 35 μm thick, 99.95%). The produced graphene on the copper foil of dimensions 20 mm × 35 mm was coated with PMMA solution in anisole (495 PMMA, Microchem) to produce PMMA solid films via spin coating [32]. The spinning conditions and the solution concentrations, listed in Table 1, were optimized accordingly to produce the desired thickness of the Gr/PMMA films. Then, the sample was allowed to float on a 0.15 M aqueous etchant solution of ammonium persulphate (APS) to etch away the copper substrate. After copper etching, the floating Gr/PMMA membrane was thoroughly rinsed with deionized-double distilled water until the APS solution was fully replaced. Then the floating film was deposited on another Gr/PMMA layer on a copper foil (which serves as the sacrificial substrate for the repetitive film depositions) by removing the water, as detailed elsewhere [32]. The deposited film was dried at 40 °C for several hours, and then it was post-baked at 150 °C for 5 min on a hot plate. This procedure was repeated until the desired number of layers was achieved. To separate the Gr/PMMA stacks from the copper substrate, a similar APS solution was used as described above. We prepared three samples with different graphene volume fractions, which are labeled as Gr/PMMA(0.2), Gr/PMMA(3.3), and Gr/PMMA(5.1), respectively. The number of stacking periods and the corresponding volume fraction of graphene are shown in Table 1.

Table 1. Structural characteristics of the three Gr/PMMA Bragg stacks

| Specimen code | Number of PMMA layers | Thickness of one PMMA layer (nm) | Solution wt% | RPM | Volume fraction of graphene (%) |
|---|---|---|---|---|---|
| Gr/PMMA(0.2) | 6 | 1500 | 11 | 3000 | 0.02 |
| Gr/PMMA(3.3) | 15 | 100 | 2 | 1000 | 0.33 |
| Gr/PMMA(5.1) | 25 | 66 | 2 | 2000 | 0.51 |

## 2.2 Brillouin Light Spectroscopy (BLS)

It is well known the elastic modulus is coupled with the acoustic sound velocity near the Γ point in the first Brillouin zone [33, 34]. In this study, we applied BLS to determine the sound velocity of the Gr/PMMA Bragg stacks. In a typical BLS experiment, the phonon wave vector $\mathbf{q} = \mathbf{k}_s - \mathbf{k}_i$, where $\mathbf{k}_s$ and $\mathbf{k}_i$ is the wave vector of the scattered and incident light, respectively. A laser with a wavelength of 532 nm was utilized to probe the phonons, and the scattered light was detected by a six-pass tandem Fabry-Perot interferometer. The longitudinal acoustic phonon (LA) can be observed in the VV spectra, where V denotes a vertical polarization with respect to the scattering plane defined by $\mathbf{k}_i$ and $\mathbf{k}_s$. Using a VV polarization configuration, we conducted measurements in the transmission (with a scattering angle, $\theta = 90°$) and backscattering geometries to detect the LA phonon mode in the in-plane (parallel to the sample films) and cross-plane (normal to the sample films) directions, respectively. Because of light absorption [35], graphene can convert part of the incident light energy to heat, which can lead to melting of PMMA and even burning of the stacks. To avoid any hotspot, we installed a filter in front of the specimens to control the incident power, which was measured by an optical power meter (Newport Model 1961-C).

## 2.3 Tensile test

Tensile tests were performed on a micro-tensile tester equipped with a 5 N load cell (MT-200, Deben UK Ltd, Woolpit, UK). The specimens were strips of dimensions of 35 mm × 1 mm in length and width, respectively. The thickness of the films was evaluated using a digital micro-meter with a resolution of 0.1 μm (Mitutoyo, Japan). The mean film thickness was calculated based on 10 measurements within the gauge length area. During the tensile test, the two ends of the strip were fixed to a 25 mm × 25 mm paper frame by using a two-part cold curing epoxy resin. A paper frame was utilized to prevent any early failure within the gripping zone. The tests were carried out at room temperature at a strain rate of 0.008 min$^{-1}$. The stress and strain values were extracted from the recorded load and displacement raw data. The estimation of the Young's modulus was obtained by the mean values of at least 10 samples for each graphene content, and the experimental errors are the deviation from the mean values. The analysis was performed by using linear regression of the initial linear part of the stress-strain curves.

## 2.4 Raman characterization

Raman mapping was performed on an area of 50 μm × 50 μm by acquiring spectra at steps of 2 μm. A Renishaw Invia Raman Spectrometer with 2400 & 1200 grooves/mm grating for the 514 nm laser excitation and a 100× lens with a N.A. of 0.85 and W.D. of 1.0 was used to evaluate the quality of the Gr on macroscale nanolaminates. The laser power was kept below 1 mW to avoid overheating the specimens. All Raman peaks were fitted with Lorentzian functions, and the spectroscopic parameters (peak position and FWHM) were recorded at each position on the film.

## 3. Results

### 3.1 Characterization of the Gr/PMMA stacks

Figure 1(a) shows the AFM image of the investigated CVD graphene, which was placed on a Si/SiO$_2$ wafer surface by using the lift-off/float-on deposition process (Figure S1). The wrinkles introduced by the synthesis process can be clearly seen in figure S5 by the AFM line scan, and correspond to sharp elevations of maximum height of 6 nm and an average FWHM of ca. 30 nm. The stacks were then fabricated by the wet transfer method, and the corresponding SEM image of Gr/PMMA(0.2) is shown in figure 1(b). Raman spectroscopy and x-ray diffraction pattern (XRD) were employed to evaluate the quality of graphene layers during the sample preparation. In figure 1(c), two representative spectra from single and multi Gr/PMMA layers are presented. It is evident that the G and 2D peak intensities for multi Gr/PMMA layers are higher due to the contribution of multiple graphene layers to the spectrum. As estimated from the Raman wavenumber contours in figure S2, after the first deposition, the 2D peak was found to be slightly blue shifted due to the compressive fields generated (around 0.04%, figure S3(a)) during the polymer thin film preparation [36]. The other spectroscopic characteristics of graphene spectrum correspond to typical values obtained from a single layer CVD graphene with a FWHM of ~ 33 cm$^{-1}$ for the 2D peak and an intensity ratio, I(2D)/I(G), of ~ 2.3 (figure S3(a)) [37, 38]. Raman spectra were acquired during each deposition, and only minor spectral changes were observed vis-a-vis the first deposition. In figure S3, the contours of 2D peak, the FWHM of 2D peak, and the I(2D)/I(G) ratios are presented. A small broadening to the 2D peak may be attributed to expected slight wavenumber variations between each graphene layer within the sampling volume [39]. According to a previous study, both G and 2D peaks undergo red-shift and their FWMHs can be broadened by increasing the average wrinkle height and their FWHMs as well [40]. The nearly unchangeable G and 2D peak positions with their FWHMs of Gr/PMMA stacks in our case indicate that no residual stresses are developed during the wet transfer/solidification process. Figure 1(d) displays the x-ray diffraction (XRD) pattern of Gr/PMMA(0.2) in the 2θ ranging from 10 to 70 degrees, which apparently shows the characteristic peaks for both graphene and PMMA. Known as an amorphous polymer, PMMA has the three broad peaks at 2θ = 13.8°, 31.8° and 40°, in which the first reflects the ordered packing of polymer chains, and the second denotes the ordering inside, respectively [41]. The peak at 2θ = 24° denotes the (200) face of graphene and the corresponding d-spacing is 3.7 Å. Upon multiple depositions of Gr/PMMA layers, a slight increase of the I(D)/I(G) ratio vis-à-vis a single Gr/PMMA layer is observed; revealing a minor inducement of structural defects by the sequential assembly of the Gr/PMMA hybrid Bragg stacks (figure S4). It is clear from the contours (figure S2-S3) and the XRD pattern that the graphene quality was not seriously affected by the sample preparation since the spectral characteristics are similar to those obtained for the single layer.

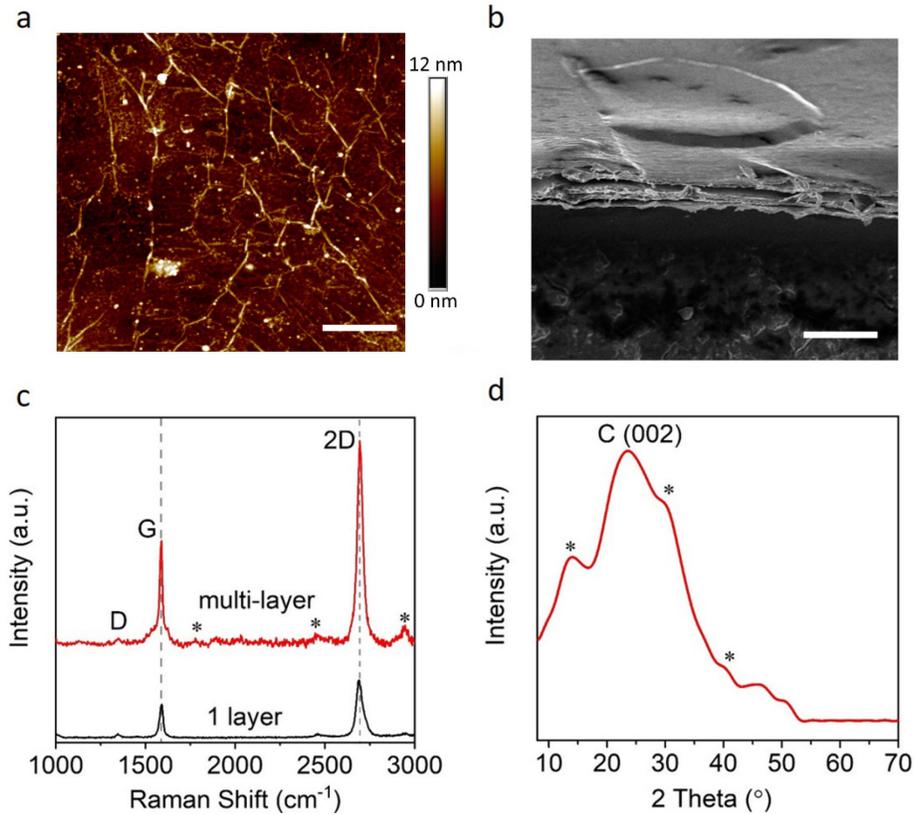

**Figure 1. Characterization of the Gr/PMMA multilayer stacks.** (a) AFM image of the CVD graphene with a maximum wrinkle height of ~ 6 nm. The scale bar is 740 nm. (b) Side view of Gr/PMMA(0.2) under SEM where the lateral structure can be seen. The scale bar is 100 μm. (c) Representative Raman spectra collected from a single (bottom) and multiple (top) graphene/PMMA layers. Asterisks mark the spectroscopic features of PMMA. The vertical dashed lines mark the robust peak during the wet transfer (i.e., lift-on and float-off) process [42]. (d) XRD pattern for Gr/PMMA(0.2). The highest peak denotes the characteristic peak of graphene while the asterisks mark PMMA peaks.

### 3.2 Elastic modulus of CVD graphene from BLS

Figure 2 shows exemplary polarized BLS spectra (anti-Stokes side) of Gr/PMMA(0.2) recorded at a constant $q$ = 0.0167 nm$^{-1}$; the corresponding BLS spectra for pure PMMA, Gr/PMMA(3.3), and Gr/PMMA(5.1) are shown in figure S6-S8. The peak position of the spectra defines the frequency of the longitudinal phonon referring to the effective medium acoustic phonon in the Gr/PMMA Bragg stack. As a dispersive medium, PMMA exhibits fast segmental dynamics at elevated temperatures well above its glass transition temperature (around 380 K). Under ambient conditions, the sound propagation is dissipation free [43]. Conversely, the sound velocity in Gr is nearly independent of temperature in the range of 300-500 K [44], which covers the local temperatures in our experiments. The relatively stable sound velocity of graphene allows us to measure it by probing the Gr/PMMA stacks. Therefore, the red shift of the LA frequency in the two scattering geometries is simply attributed to the decreasing sound velocity of PMMA with increasing laser power.

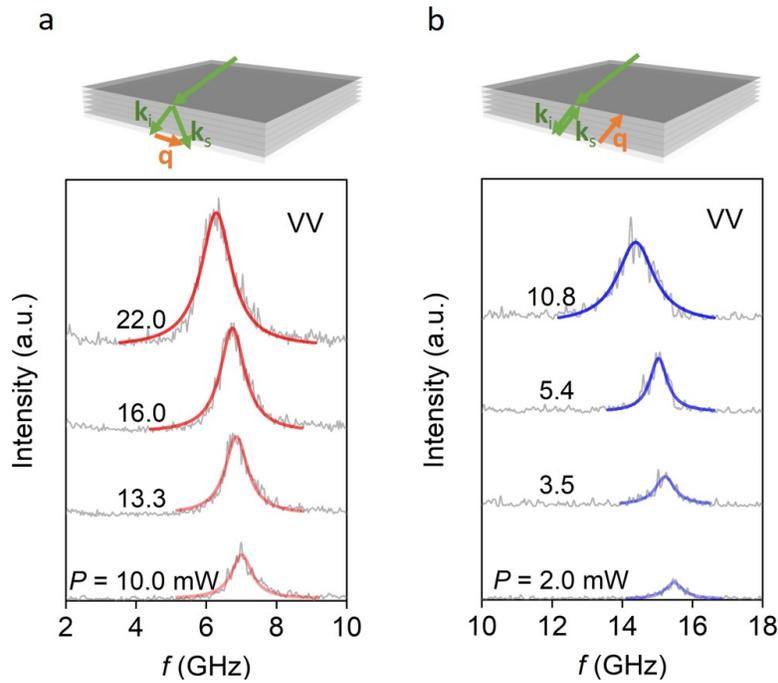

**Figure 2. Brillouin light scattering spectra of Gr/PMMA(0.2).** The exemplary spectra recorded in the (a) transmission and (b) backscattering geometries are represented by single Lorentzian peaks (red and blue curves). VV denotes a vertical polarization configuration of the incident and scattered light with respect to the scattering plane of $k_i$ and $k_s$ (schemes in the top panels). At a higher incident power, a clear red shift of the peak position is observed in both geometries. In the schematics, the thermally excited phonons are probed by the incident light with a wave vector $k_i$, and the scattered light has a wave vector $k_s$. The phonon wave vector is $q = k_s - k_i$. The grey and black sheets denote PMMA and CVD graphene layers, respectively.

Figure 3(a) & (b) show the power-dependent longitudinal sound velocity in the in-plane ($c_{L,\parallel}$) and cross-plane ($c_{L,\perp}$) direction, respectively. For Gr/PMMA(5.1), $c_{L,\parallel} = \frac{2\pi f}{q}$, are 3070 and 2434 m s$^{-1}$ at $P$ = 1.66 and 6.85 mW, respectively. When the incident power exceeds a threshold, a sharper decrease of $c_{L,\parallel}$ is observed. By linear fitting of the data before and after the sharp decrease, a kink is clearly seen. The kink is rationalized by the glass transition of PMMA [45], at which the phonon vibration undergoes an abrupt softening. At much higher temperatures, the velocity corresponds to the adiabatic zero frequency sound velocity, $c_0$, while at much lower temperatures it corresponds to the solid-like or 'infinite' frequency sound velocity, $c_\infty$ [44]. Thus, the decrease rates of the $c_{L,\parallel}$ in Gr/PMMA with respect to temperature are different in the glassy and rubbery states. The kink for Gr/PMMA(0.2) appears at $P$ = 8 mW, whereas the kink positions for the other two samples are at much lower power levels, owing to the larger number of graphene layers (i.e., more light absorption in Gr/PMMA(3.3) and Gr/PMMA(5.1) than Gr/PMMA(0.2)); the sound velocity of the fully transparent PMMA is expectedly robust to the laser power variation. Another strong justification of the kink position is the linewidth of BLS peaks, *i.e.*, the ω-spread of the phonons, related to the phonon mean free path [46]. An apparent increase in the linewidth is observable after the kink (figure S9). The intercept (at $P$ = 0 mW) of the fitted line before the kink defines the values of $c_{L,\parallel}$, $c_{L,\perp}$ at room temperature (i.e., without heating).

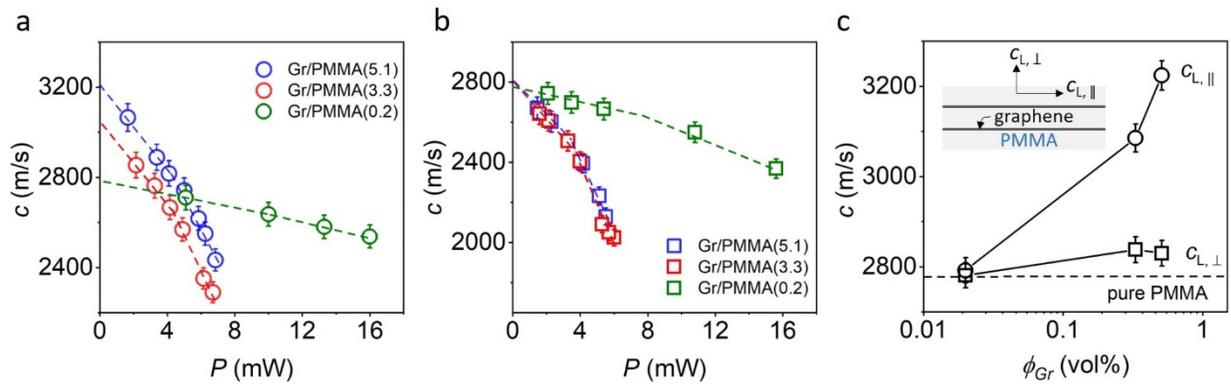

**Figure 3. Longitudinal sound velocity, $c_L$ of the CVD graphene/PMMA hybrid Bragg stacks.** The variation of $c_L$, in-plane, and cross-plane phonon propagation with the incident laser power are shown in (a) and (b), respectively. The PMMA is in the glassy (rubbery) state at laser powers lower (higher) than the value at the kink. The intercepts (at $P = 0$ mW) of the linear fits to the data denote the sound velocities of the samples at 295K. The dashed lines are linear fits to the data. (c) The effective in-plane $c_{L,\parallel}$ and cross-plane $c_{L,\perp}$ as shown schematically in the inset, are plotted as a function of the graphene volume fraction percentage. The circles (squares) data symbols denote the in-plane (cross-plane) $c_L$ and the two lines are guide to the eye. The black dashed line represents the sound velocity of pure PMMA.

### 3.3 Band structure calculation

Figure 3(c) shows the sound velocity $c_L$ as a function of the graphene volume fraction. The $c_{L,\parallel}$ increases from $2790 \pm 50$ to $3224 \pm 60$ m s$^{-1}$ as the graphene volume fraction increases from 0.02 vol% to 0.51 vol%. The $c_{L,\parallel}$ of Gr/PMMA(0.2) is very close to the value of pure PMMA (2784 m s$^{-1}$) due to the extremely low graphene volume fraction. The $c_{L,\perp}$ for all specimens remains nearly the same to that of the PMMA as the direction normal to the Bragg stack [29, 30] is dominated by the polymer. From the BLS experiments, we directly obtained the longitudinal sound velocities of PMMA and the effective longitudinal sound velocities of the Gr/PMMA stacks. To determine the elastic moduli of the CVD graphene without the assumption of effective medium approximations [29], we conducted phononic band structure calculations by using the COMSOL Multiphysics package. Because of symmetry, we conducted two-dimensional (2D) simulations. The simulation domain consists of one layer of graphene and one layer of PMMA, with periodic boundary conditions applied on all four sides. The relevant parameters are listed in Table 2. Specifically, we assumed the Poisson's ratio of graphene to be 0.17 [47] and considered the Young's modulus of the CVD graphene as an adjustable parameter. The elastic properties of the relatively thick PMMA layers were assumed to be the bulk PMMA values (Young's modulus = 6.272 GPa, Poisson's ratio = 0.333, based on our BLS measurements). By matching the measured effective longitudinal sound velocities of the Gr/PMMA stacks with the corresponding calculated values in the long wavelength limit, we determined the Young's modulus of the investigated CVD graphene to be $680 \pm 16$ GPa. Typical phononic band structures are shown in figure 4(a) and 4(b) for Gr/PMMA(3.3) in the cross-plane and in-plane directions, respectively. The periodicity-induced band folding is clearly seen in figure 4(a). For the band structure in the in-plane direction, we considered a small domain width of 1 nm to avoid band folding in the width direction. The slopes of the lowest two branches give the effective longitudinal and transverse sound velocities of the Gr/PMMA(3.3) stack, respectively. Figure 4(c) shows effective longitudinal and transverse sound velocities of the Gr/PMMA stacks as a function of the graphene volume fraction. It is interesting to note that as the graphene volume fraction increases, $c_{L,\parallel}$ increases, whereas $c_{L,\perp}$ and $c_{T,\parallel}$ ($\approx c_{T,\perp}$) remain nearly constant. In the limit of zero graphene volume fraction, $c_{L,\parallel} \approx c_{L,\perp}$ expectedly erasing the elastic anisotropy, $(c_{L,\parallel} / c_{L,\perp})^2$, of the PMMA rich Bragg stack, and $c_L$ assumes the sound velocity of pure PMMA (black dashed line in figure 3(c)). Assuming an in-plane isotropy of graphene, we further obtained the shear modulus of graphene to be $G = E/[2(1 + v)] = 290 \pm 10$ GPa, which is in good agreement with the reported value (280 GPa at 0.4K) for CVD graphene on silicon mechanical oscillator [48].

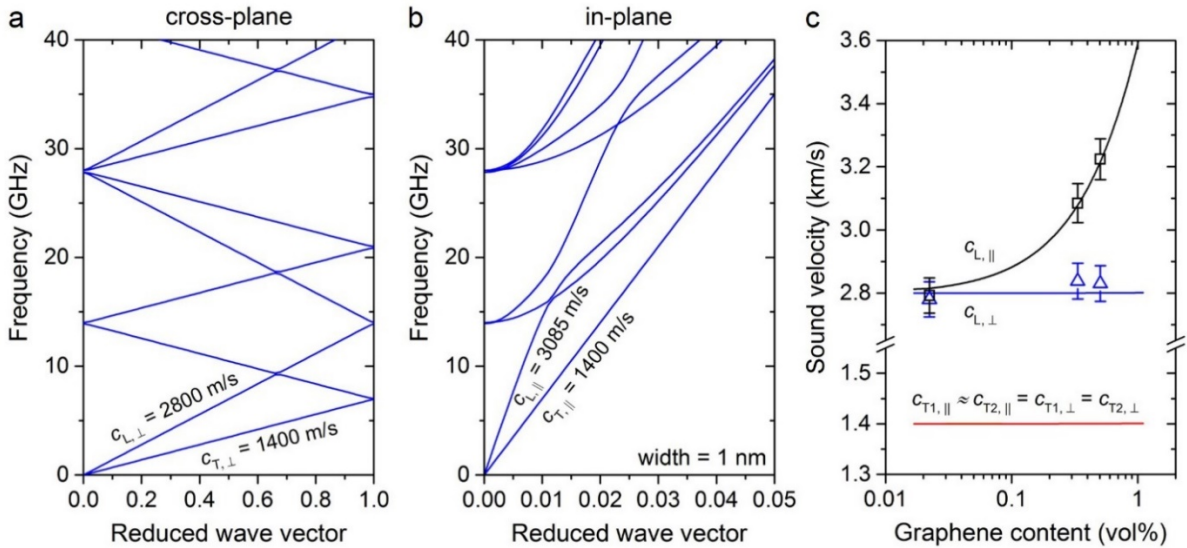

**Figure 4. Calculated phononic band structure and effective sound velocities in Gr/PMMA Bragg stacks.** The phononic band structures of Gr/PMMA(3.3) in the (a) cross-plane and (b) in-plane directions. The effective sound velocities of the lowest longitudinal and transverse branches in the long wavelength limit are indicated. (c) The effective longitudinal (L) and transverse (T) sound velocities of the Gr/PMMA stacks as a function of the graphene volume fraction.

Table 2. Parameters used in the phononic band structure calculations and Young's modulus of graphene from tensile tests.

| Materials | Thickness (nm) | Width (nm) | Density (kg/m$^3$) | Young's modulus (GPa) (computed) | Poisson's ratio | Young's modulus (GPa) (tensile test) |
|---|---|---|---|---|---|---|
| Graphene | 0.335 | 1 | 2.267 | 680 ± 16 | 0.17 [47] | 817 ± 23 |
| PMMA | 3.125-100 | 1 | 1.189 | 6.2±0.2 | 0.333 | 1.8 |

### 3.4 Uniaxial tensile modulus

Similar to our previous work [32], uniaxial tensile tests were conducted to evaluate the effect of CVD graphene on the elastic moduli of the Gr/PMMA hybrid Bragg stacks. In Figure 5(a) exemplary stress-strain curves are presented for each volume fraction. The Young's modulus of each Gr/PMMA specimen was estimated from the linear part (0.1-0.3% strain) of the stress-strain curves. For similar systems, it has been shown that even for a low thickness ratio of the two components, e.g., 0.044 vol% graphene, the modulus of elasticity of the composite shows a significant increase of 25% [32]. For the Gr/PMMA(3.3), with the thickness of a PMMA layer being 100 nm, the Young's modulus increases to 4.4 ± 0.4 GPa, 150% higher than that of the bulk PMMA. The subsequent reduction in thickness of the matrix to 65 nm (Gr/PMMA(5.1)) yields an even larger increase of Young's modulus to 5.9 ± 0.4 GPa (a 250% increase). By using a simple rule-of-mixtures: $E_{Gr/PMMA} = E_m \phi_{PMMA} + E_f \phi_{Gr}$, where $E_{Gr/PMMA}$, $E_{PMMA}$, and $E_{Gr}$ are the Young's moduli of the composite, PMMA, and graphene, respectively. $\phi_{PMMA}$ and $\phi_{Gr} = 1 - \phi_{PMMA}$ are the volume fractions of the PMMA and graphene, respectively. From a linear least-squares-fit (presented in SI) of the data in figure 5(b) extrapolating to 100% graphene volume fraction, the Young's modulus of the graphene was estimated to 817±23 GPa.

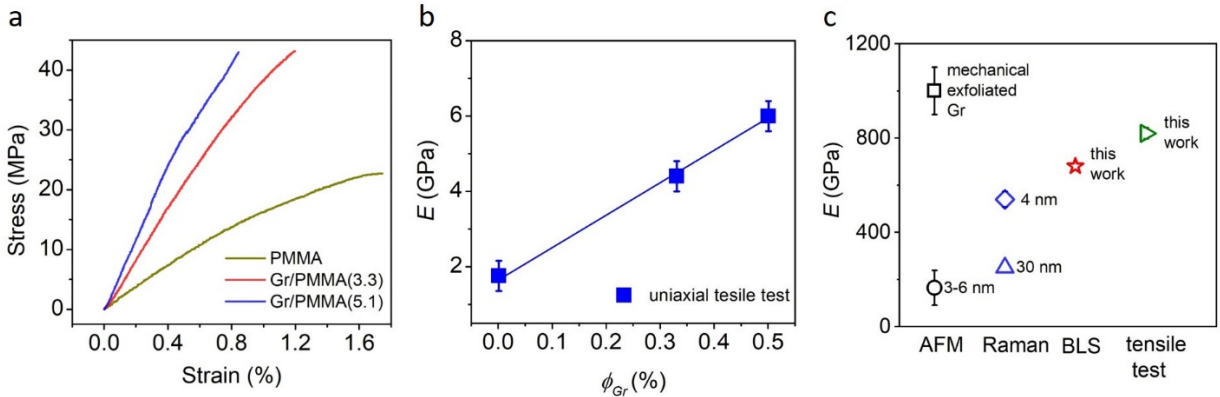

**Figure 5.** (a) Representative stress-strain curves of pure PMMA, Gr/PMMA(3.3) and Gr/PMMA(5.1). (b) The Young's modulus in relation to the graphene volume fraction as obtained from tensile measurements. The solid is a linear-squares-fit to the experimental data of a slope of 8.15 GPa. (c) Experimental Young's moduli in CVD graphene of different wrinkle height by different techniques. Mechanically exfoliated graphene: ref [3]; CVD graphene having a wrinkle height of 3-6 nm: ref [25]; 4-8 nm: ref [49]; 30 nm: ref [24].

## 4. Discussion and concluding remarks

We summarize the experimental Young's modulus of wrinkled graphene, as deduced from different experimental techniques, in figure 5(c) to visualize the comparison among them. AFM nanoindentation of CVD graphene with wrinkle height between 3-6 nm yielded to a Young's modulus of $167 \pm 74$ GPa [25], while a similar indentation technique with a gating voltage led to $E = 100 \pm 86$ GPa for CVD graphene with wrinkle heights of 3 nm [18]. However, nanoindentation itself can cause a large deformation in the out-of-plane direction, which may deteriorate the intrinsic stiffness of graphene. Yet, the strain distribution in wrinkled graphene is non-uniform, introducing large errors in the measured stiffness. Raman spectroscopy, which estimates the Young's modulus of a 2D material from the rate of the frequency shift of the characteristic peaks under strain, reported a Young's modulus of 250 GPa [24] for graphene with an average wrinkle height of ca. 30 nm. A more comprehensive summary of the Young's modulus of graphene, estimated by various experimental techniques, can be found in Table. S1.

The uniaxial tensile test is widely applied to measure the mechanical characteristics of the Bragg stacks such as the Young's modulus and the fracture stress/strain. For the specimens tested here the Young's modulus of CVD graphene was estimated to be 817 ± 23 GPa about 20% higher than 680 ± 16 GPa obtained from BLS. The main reason for this discrepancy is attributed to the straightening of the wrinkles upon the imposition of a tensile stress which results in higher Young's modulus than that estimated by the BLS method at zero strain [49]. A possible source of error is also the linear extrapolation invoked in figure 5(b) which may not be holding to 100% volume fraction. We point out that in this work, the wrinkles in the CVD graphene were mainly introduced upon cooling of the CVD graphene on the copper substrate and the subsequent transfer process.[50]

Unlike the methods mentioned above, BLS allows probing thermally excited phonons in a non-destructive manner. It is well-known that elastic properties are coupled with the acoustic sound velocities; therefore, accurate elastic moduli can be obtained by probing the effective sound velocity of the hybrid Bragg stacks. Our experiment provides a method for studying the elastic modulus of other two-dimensional materials, such as reduced graphene oxide (rGO) and graphene oxide (GO), which are commonly used fillers to reinforce the elastic properties of polymer-based composites. However, the wrinkle-induced softening effect in these materials has not yet been investigated. As a mature technique, the wet transfer can be easily employed to these 2D materials, and by probing the GO/PMMA or rGO/PMMA hybrid Bragg stacks using BLS, the longitudinal sound velocity can then be obtained. The elastic moduli can be subsequently calculated with the known Poisson's ratios. These results can benefit an improved understanding of the elastic moduli of 2D materials and pave the way to accurately manipulating the elastic moduli of the composites.

Concluding, we report here a new approach to accessing the elastic modulus of the CVD graphene. By probing the effective phonon propagation of Gr/PMMA hybrid Bragg stacks using BLS and combining the phononic band structure calculations, we determined the Young's and shear moduli of the CVD graphene with a maximum height of 6 nm to be 680 ± 16 and 290 ± 10 GPa, respectively. In probing the thermal phonons, no strain was introduced, eliminating the possibility of strain hardening upon loading. This concept can be extended to determining the wrinkle-induced softening effect of other two-dimensional materials.

**Credit authorship contribution statement**

**Bohai Liu:** Investigation, Formal analysis, Visualization, Writing - original Draft, Funding acquisition. **Christos Pavlou:** Resources, Formal analysis, Investigation, Writing - original Draft. **Zuyuan Wang:** Software, Writing - original Draft. **Yu Cang:** Methodology, Formal analysis. **Costas Galiotis:** Writing - review & editing, Supervision, Funding acquisition. **George Fytas:** Writing - review & editing, Supervision, Funding acquisition.

**Acknowledgments**

B.L. gratefully thanks the financial support from China Scholarships Council (No. 201906260224). Z.W. and G.F. acknowledge the financial support by ERC AdG SmartPhon (Grant No. 694977). Y.C. acknowledge the financial support by Shanghai Pujiang Program (Grant No. 20PJ1413800). C.G. and C.P. acknowledge the financial support of "Graphene Core 3, GA: 881603 – Graphene-based disruptive technologies", which is implemented under the EU-Horizon 2020 Research & Innovation Action. C.G. and C.P. are also thankful to Dr. Anastasios Manikas for assisting with the Raman measurements, Dr. Giovanna-Maria Pastore-Carbone for assisting with the nanocomposite fabrication and Dr. George Trakakis for providing the CVD graphene samples. The authors declare that they have no known competing financial interests or personal relationships that could have appeared to influence the work reported in this paper.

**ReferenceUncategorized References**

[1] A.A. Balandin, S. Ghosh, W. Bao, I. Calizo, D. Teweldebrhan, F. Miao, C.N. Lau, Superior Thermal Conductivity of Single-Layer Graphene, Nano Letters 8(3) (2008) 902-907.
[2] K.S. Kim, Y. Zhao, H. Jang, S.Y. Lee, J.M. Kim, K.S. Kim, J.H. Ahn, P. Kim, J.Y. Choi, B.H. Hong, Large-scale pattern growth of graphene films for stretchable transparent electrodes, Nature 457(7230) (2009) 706-10.

[3] C. Lee, X. Wei, J.W. Kysar, J. Hone, Measurement of the Elastic Properties and Intrinsic Strength of Monolayer Graphene, Science 321(5887) (2008) 385.
[4] K.S. Novoselov, A.K. Geim, S.V. Morozov, D. Jiang, Y. Zhang, S.V. Dubonos, I.V. Grigorieva, A.A. Firsov, Electric Field Effect in Atomically Thin Carbon Films, Science 306(5696) (2004) 666.
[5] F. Liu, P. Ming, J. Li, Ab initiocalculation of ideal strength and phonon instability of graphene under tension, Physical Review B 76(6) (2007).
[6] T. Kuilla, S. Bhadra, D. Yao, N.H. Kim, S. Bose, J.H. Lee, Recent advances in graphene based polymer composites, Progress in Polymer Science 35(11) (2010) 1350-1375.
[7] D.G. Papageorgiou, I.A. Kinloch, R.J. Young, Mechanical properties of graphene and graphene-based nanocomposites, Progress in Materials Science 90 (2017) 75-127.
[8] I. Vlassiouk, G. Polizos, R. Cooper, I. Ivanov, J.K. Keum, F. Paulauskas, P. Datskos, S. Smirnov, Strong and electrically conductive graphene-based composite fibers and laminates, ACS Appl Mater Interfaces 7(20) (2015) 10702-9.
[9] K. Liu, J. Wu, Mechanical properties of two-dimensional materials and heterostructures, Journal of Materials Research 31(7) (2015) 832-844.
[10] Q. Yu, J. Lian, S. Siriponglert, H. Li, Y.P. Chen, S.-S. Pei, Graphene segregated on Ni surfaces and transferred to insulators, Applied Physics Letters 93(11) (2008).
[11] K.S. Novoselov, V.I. Fal′ko, L. Colombo, P.R. Gellert, M.G. Schwab, K. Kim, A roadmap for graphene, Nature 490(7419) (2012) 192-200.
[12] X. Li, W. Cai, J. An, S. Kim, J. Nah, D. Yang, R. Piner, A. Velamakanni, I. Jung, E. Tutuc, S.K. Banerjee, L. Colombo, R.S. Ruoff, Large-Area Synthesis of High-Quality and Uniform Graphene Films on Copper Foils, Science 324(5932) (2009) 1312.
[13] Y. Zhang, L. Zhang, C. Zhou, Review of Chemical Vapor Deposition of Graphene and Related Applications, Accounts of Chemical Research 46(10) (2013) 2329-2339.
[14] Y. Kim, J. Lee, M.S. Yeom, J.W. Shin, H. Kim, Y. Cui, J.W. Kysar, J. Hone, Y. Jung, S. Jeon, S.M. Han, Strengthening effect of single-atomic-layer graphene in metal-graphene nanolayered composites, Nat Commun 4 (2013) 2114.
[15] P.Y. Huang, C.S. Ruiz-Vargas, A.M. van der Zande, W.S. Whitney, M.P. Levendorf, J.W. Kevek, S. Garg, J.S. Alden, C.J. Hustedt, Y. Zhu, J. Park, P.L. McEuen, D.A. Muller, Grains and grain boundaries in single-layer graphene atomic patchwork quilts, Nature 469(7330) (2011) 389.
[16] A. Shekhawat, R.O. Ritchie, Toughness and strength of nanocrystalline graphene, Nat Commun 7 (2016) 10546.
[17] M.G. Pastore Carbone, A.C. Manikas, I. Souli, C. Pavlou, C. Galiotis, Mosaic pattern formation in exfoliated graphene by mechanical deformation, Nature Communications 10(1) (2019) 1572.
[18] R.J. Nicholl, H.J. Conley, N.V. Lavrik, I. Vlassiouk, Y.S. Puzyrev, V.P. Sreenivas, S.T. Pantelides, K.I. Bolotin, The effect of intrinsic crumpling on the mechanics of free-standing graphene, Nat Commun 6 (2015) 8789.
[19] W. Zhu, T. Low, V. Perebeinos, A.A. Bol, Y. Zhu, H. Yan, J. Tersoff, P. Avouris, Structure and electronic transport in graphene wrinkles, Nano Lett 12(7) (2012) 3431-6.
[20] X. Li, Y. Zhu, W. Cai, M. Borysiak, B. Han, D. Chen, R.D. Piner, L. Colombo, R.S. Ruoff, Transfer of Large-Area Graphene Films for High-Performance Transparent Conductive Electrodes, Nano Letters 9(12) (2009) 4359-4363.
[21] N. Liu, Z. Pan, L. Fu, C. Zhang, B. Dai, Z. Liu, The origin of wrinkles on transferred graphene, Nano Research 4(10) (2011) 996.
[22] T. Jiang, Z. Wang, X. Ruan, Y. Zhu, Equi-biaxial compressive strain in graphene: Grüneisen parameter and buckling ridges, 2D Materials 6(1) (2018) 015026.
[23] X. Shen, X. Lin, N. Yousefi, J. Jia, J.-K. Kim, Wrinkling in graphene sheets and graphene oxide papers, Carbon 66 (2014) 84-92.
[24] Z. Li, I.A. Kinloch, R.J. Young, K.S. Novoselov, G. Anagnostopoulos, J. Parthenios, C. Galiotis, K. Papagelis, C.-Y. Lu, L. Britnell, Deformation of Wrinkled Graphene, ACS Nano 9(4) (2015) 3917-3925.

[25] C.S. Ruiz-Vargas, H.L. Zhuang, P.Y. Huang, A.M. van der Zande, S. Garg, P.L. McEuen, D.A. Muller, R.G. Hennig, J. Park, Softened Elastic Response and Unzipping in Chemical Vapor Deposition Graphene Membranes, Nano Letters 11(6) (2011) 2259-2263.
[26] A.C. Ferrari, D.M. Basko, Raman spectroscopy as a versatile tool for studying the properties of graphene, Nat Nanotechnol 8(4) (2013) 235-46.
[27] O. Frank, G. Tsoukleri, I. Riaz, K. Papagelis, J. Parthenios, A.C. Ferrari, A.K. Geim, K.S. Novoselov, C. Galiotis, Development of a universal stress sensor for graphene and carbon fibres, Nature Communications 2(1) (2011) 255.
[28] C. Androulidakis, G. Tsoukleri, N. Koutroumanis, G. Gkikas, P. Pappas, J. Parthenios, K. Papagelis, C. Galiotis, Experimentally derived axial stress–strain relations for two-dimensional materials such as monolayer graphene, Carbon 81 (2015) 322-328.
[29] D. Schneider, F. Liaqat, E.H. El Boudouti, Y. El Hassouani, B. Djafari-Rouhani, W. Tremel, H.-J. Butt, G. Fytas, Engineering the Hypersonic Phononic Band Gap of Hybrid Bragg Stacks, Nano Letters 12(6) (2012) 3101-3108.
[30] Z. Wang, K. Rolle, T. Schilling, P. Hummel, A. Philipp, B.A.F. Kopera, A.M. Lechner, M. Retsch, J. Breu, G. Fytas, Tunable Thermoelastic Anisotropy in Hybrid Bragg Stacks with Extreme Polymer Confinement, Angewandte Chemie International Edition 59(3) (2020) 1286-1294.
[31] M. Hesami, A. Gueddida, N. Gomopoulos, H.S. Dehsari, K. Asadi, S. Rudykh, H.J. Butt, B. Djafari-Rouhani, G. Fytas, Elastic wave propagation in smooth and wrinkled stratified polymer films, Nanotechnology 30(4) (2019) 045709.
[32] C. Pavlou, M.G. Pastore Carbone, A. C. Manikas, G. Trakakis, C. Koral, G. Papari, A. Andreone, C. Galiotis, Record EMI shielding behaviour of thin graphene/ PMMA nanolaminates in the THz range, under revision.
[33] X. Cong, Q.-Q. Li, X. Zhang, M.-L. Lin, J.-B. Wu, X.-L. Liu, P. Venezuela, P.-H. Tan, Probing the acoustic phonon dispersion and sound velocity of graphene by Raman spectroscopy, Carbon 149 (2019) 19-24.
[34] Z.K. Wang, H.S. Lim, S.C. Ng, B. Özyilmaz, M.H. Kuok, Brillouin scattering study of low-frequency bulk acoustic phonons in multilayer graphene, Carbon 46(15) (2008) 2133-2136.
[35] R.R. Nair, P. Blake, A.N. Grigorenko, K.S. Novoselov, T.J. Booth, T. Stauber, N.M.R. Peres, A.K. Geim, Fine Structure Constant Defines Visual Transparency of Graphene, Science 320(5881) (2008) 1308.
[36] C. Androulidakis, E.N. Koukaras, J. Parthenios, G. Kalosakas, K. Papagelis, C. Galiotis, Graphene flakes under controlled biaxial deformation, Scientific Reports 5(1) (2015) 18219.
[37] G. Anagnostopoulos, P.-N. Pappas, Z. Li, I.A. Kinloch, R.J. Young, K.S. Novoselov, C.Y. Lu, N. Pugno, J. Parthenios, C. Galiotis, K. Papagelis, Mechanical Stability of Flexible Graphene-Based Displays, ACS Applied Materials & Interfaces 8(34) (2016) 22605-22614.
[38] G. Wang, Z. Dai, L. Liu, H. Hu, Q. Dai, Z. Zhang, Tuning the Interfacial Mechanical Behaviors of Monolayer Graphene/PMMA Nanocomposites, ACS Applied Materials & Interfaces 8(34) (2016) 22554-22562.
[39] I. Vlassiouk, G. Polizos, R. Cooper, I. Ivanov, J.K. Keum, F. Paulauskas, P. Datskos, S. Smirnov, Strong and Electrically Conductive Graphene-Based Composite Fibers and Laminates, ACS Applied Materials & Interfaces 7(20) (2015) 10702-10709.
[40] Y. Zhao, X. Liu, D.Y. Lei, Y. Chai, Effects of surface roughness of Ag thin films on surface-enhanced Raman spectroscopy of graphene: spatial nonlocality and physisorption strain, Nanoscale 6(3) (2014) 1311-1317.
[41] M. Khairy, N.H. Amin, R. Kamal, Optical and kinetics of thermal decomposition of PMMA/ZnO nanocomposites, Journal of Thermal Analysis and Calorimetry 128(3) (2017) 1811-1824.
[42] C. Koral, G. Papari, M.G.P. Carbone, C. Pavlou, A. Manikas, G. Trakakis, C. Galiotis, A. Andreone, THz EMI Shielding in Graphene/PMMA Multilayers, 2019 44th International Conference on Infrared, Millimeter, and Terahertz Waves (IRMMW-THz), 2019, pp. 1-1.

[43] P. Voudouris, N. Gomopoulos, A. Le Grand, N. Hadjichristidis, G. Floudas, M.D. Ediger, G. Fytas, Does Brillouin light scattering probe the primary glass transition process at temperatures well above glass transition?, The Journal of Chemical Physics 132(7) (2010) 074906.
[44] S. Thomas, K.M. Ajith, S.U. Lee, M.C. Valsakumar, Assessment of the mechanical properties of monolayer graphene using the energy and strain-fluctuation methods, RSC Advances 8(48) (2018) 27283-27292.
[45] M. Amirkhani, A. Taschin, R. Cucini, P. Bartolini, D. Leporini, R. Torre, Polymer thermal and acoustic properties using heterodyne detected transient grating technique, Journal of Polymer Science Part B: Polymer Physics 49(9) (2011) 685-690.
[46] M. Mattarelli, M. Vassalli, S. Caponi, Relevant Length Scales in Brillouin Imaging of Biomaterials: The Interplay between Phonons Propagation and Light Focalization, ACS Photonics (2020).
[47] G. Cao, Atomistic studies of mechanical properties of graphene, Polymers 6(9) (2014) 2404-2432.
[48] X. Liu, T.H. Metcalf, J.T. Robinson, B.H. Houston, F. Scarpa, Shear modulus of monolayer graphene prepared by chemical vapor deposition, Nano Lett 12(2) (2012) 1013-7.
[49] Q.-Y. Lin, G. Jing, Y.-B. Zhou, Y.-F. Wang, J. Meng, Y.-Q. Bie, D.-P. Yu, Z.-M. Liao, Stretch-Induced Stiffness Enhancement of Graphene Grown by Chemical Vapor Deposition, ACS Nano 7(2) (2013) 1171-1177.
[50] B. Huet, J.-P. Raskin, D.W. Snyder, J.M. Redwing, Fundamental limitations in transferred CVD graphene caused by Cu catalyst surface morphology, Carbon 163 (2020) 95-104.

Supporting Information

Determination of the elastic moduli of CVD graphene by probing graphene/polymer Bragg stacks

Bohai Liu, Christos Pavlou, Zuyuan Wang, Yu Cang, Costas Galiotis and George Fytas

Contents

## 1. Optical microscope images

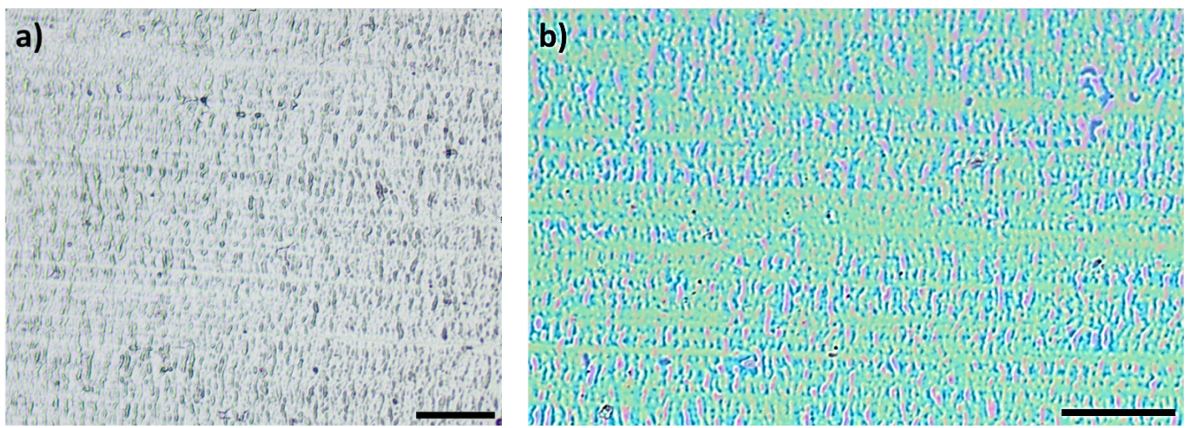

**Figure S1.** Optical microscope images of a) the surface morphology of the Cu foil after graphene growth, b) Gr/PMMA film as deposited on Si/SiO$_2$ surface (scale bar is 20um).

## 2. Raman spectroscopy

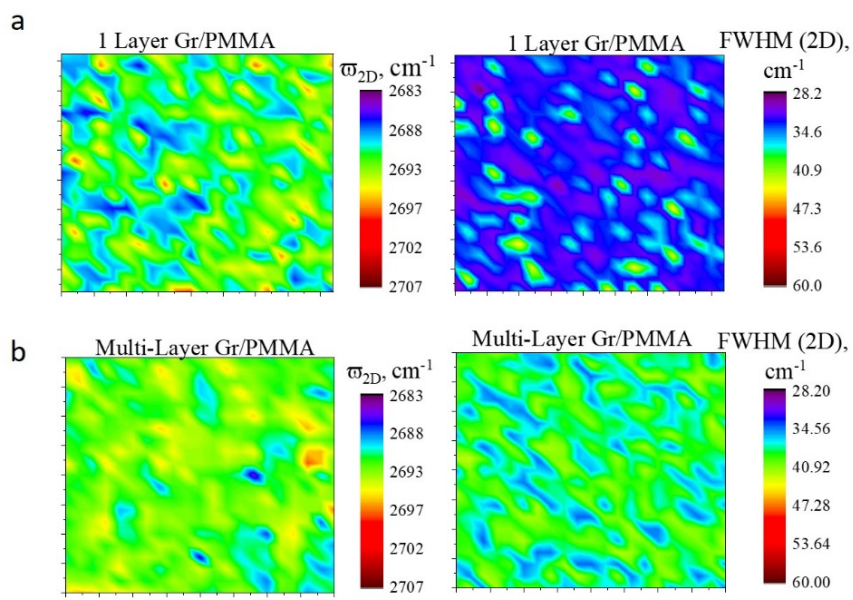

**Figure S2.** Contour maps of Pos(2D) and FWHM(2D) for (a) a single-layer Gr/PMMA film and (b) a multiple-layer Gr/PMMA film.

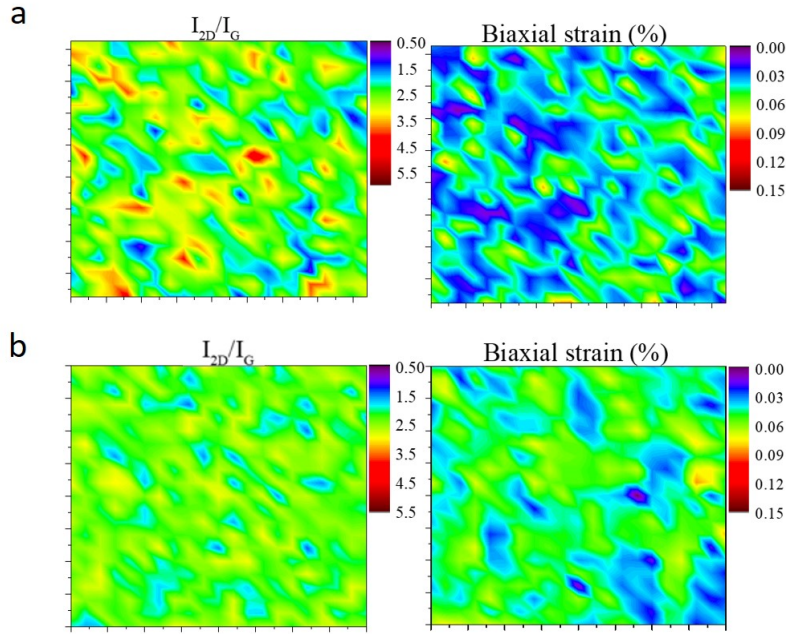

**Figure S3.** Contour maps of the I(2D)/I(G) and biaxial strain for (a) a single-layer Gr/PMMA film and (b) a multiple-layer Gr/PMMA film.

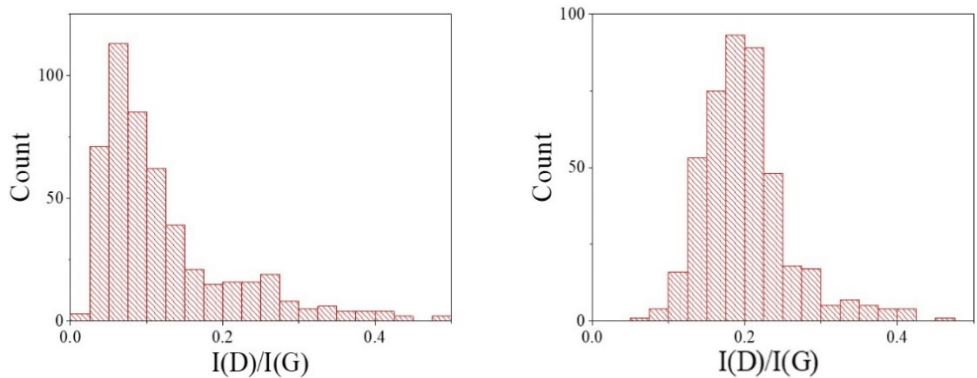

**Figure S4.** Histograms of the value of I(D)/I(G) ratios for 1Gr/PMMA (left) and multiple Gr/PMMA layers (right).

## 3. AFM characterization

The investigated CVD graphene was placed on a Si/SiO$_2$ wafer surface by using the lift-off/float-on deposition process by using an ultra-thin PMMA film as a media transfer. Then, the PMMA was removed by adopting the common procedure of immersing it in a bath of pure acetone. The speckles observed correspond to the common PMMA residues which are usually shown by this polymer removal process.

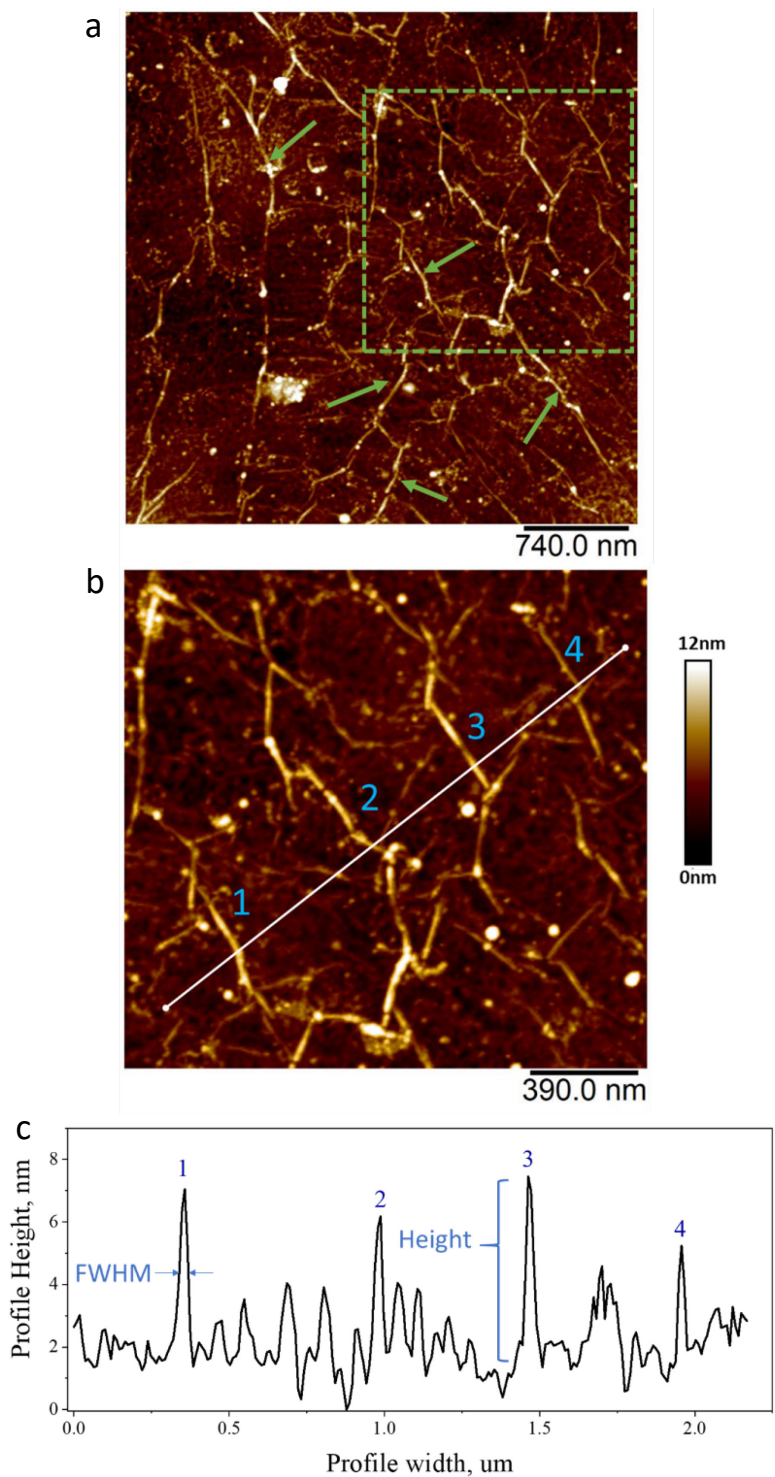

**Figure S5.** A) Wrinkles as illustrated by the AFM image (green arrows) of the CVD graphene b) Magnified image of the first image (dashed green box) and c) the corresponding line scan of the CVD graphene. Sharp elevations of a maximum height of 6 nm within experimental error can be observed.

## 3. Brillouin light spectroscopy (BLS) spectra of pure PMMA, Gr/PMMA(3.3), and Gr/PMMA(5.1)

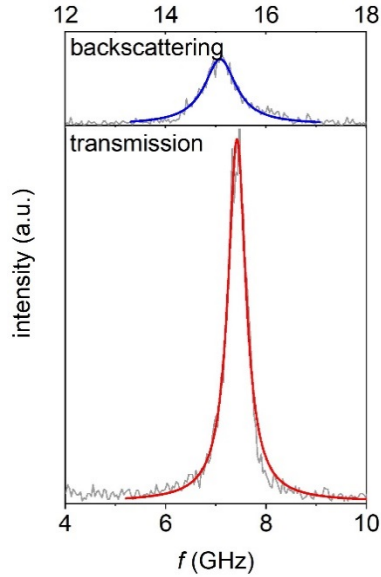

**Figure S6**. Brillouin spectra of pure PMMA thin films recorded in the transmission and backscattering geometries, which were represented by single red and blue lorentz curves, respectively.

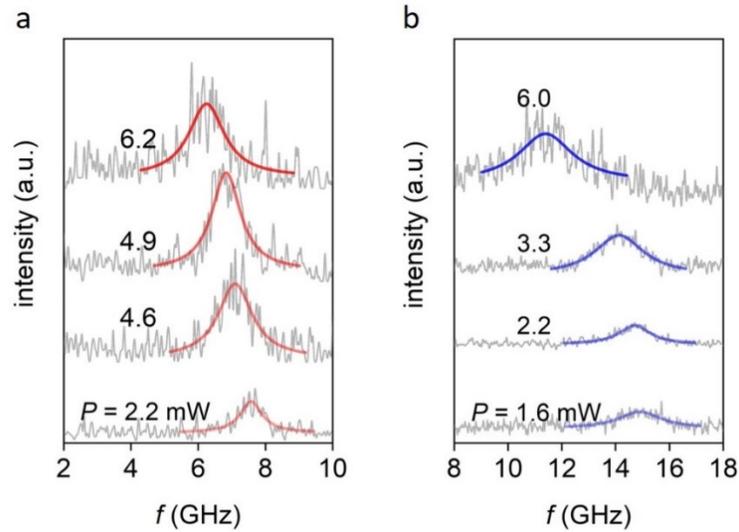

**Figure S7**. Brillouin spectra of Gr/PMMA(3.3) recorded in the (a) transmission and (b) backscattering geometries using different powers of the incident laser light. The red and blue solid lines indicate the representation of the experimental spectra by single Lorentzian lines.

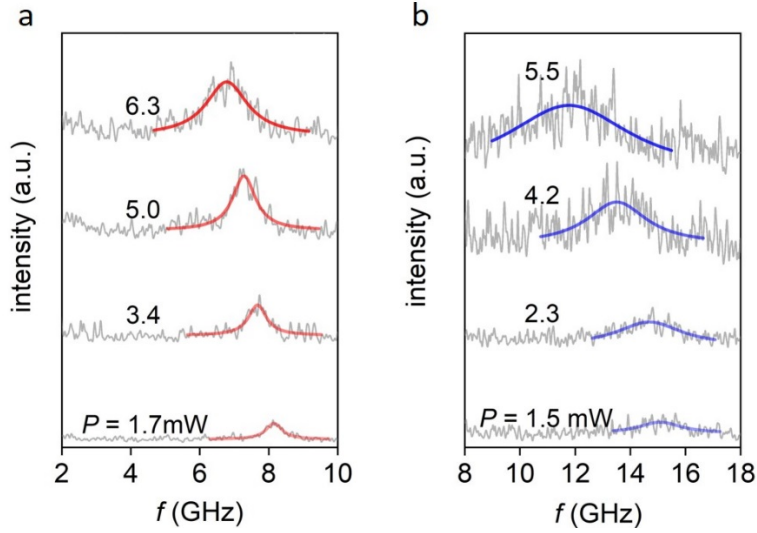

**Figure S8**. Brillouin spectra of Gr/PMMA(5.1) recorded in the (a) transmission and (b) backscattering geometries using different powers of the incident laser light. The red and blue solid lines indicate the representation of the experimental spectra by single Lorentzian lines.

### 4. The linewidths of the BLS spectra

In the low hypersonic attenuation regime ($T < T_g + 100$ K), the BLS peaks at a scattering wave vector, **q**, are well approximated by two symmetrically shifted Lorentzian curves [1] with a peak frequency $f = c/(2\pi/q)$ and a full width at half maximum, $\Gamma$. The latter defines the average phonon mean free path (MFP) [1, 2], $l = c/\Gamma$, where $c$ is the longitudinal sound velocity. Figure S6 shows the linewidths of the BLS spectra recorded in the backscattering geometry as a function of the incident laser power, $P$. A sharp increase of $\Gamma(P)$ is observed for all three Gr/PMMA stacks when the incident power exceeds a threshold (indicated by vertical lines in figure S6). In the hybrid Bragg stacks, the graphene layers convert part of the absorbed light to heat, leading to a steady state with the temperature at the laser spot higher than the ambient temperature. As the power increases, the PMMA layers undergo a glass-rubber crossover at $T_g$ accompanied by a sudden increase in the phonon attenuation ($\sim\Gamma$). Below $T_g$ (i.e., at low laser powers), the glassy PMMA displays narrow and virtually linewidths, whereas above $T_g$ (i.e., at laser powers higher than a threshold value) the phonon MFP suddenly decreases due to phonon scattering or increase of the dynamic shear viscosity, $\eta = \rho\Gamma/q^2$. The kink in the $\Gamma(P)$ plot, occurs at the same threshold incident laser power, at which the $c(P)$ (Figure 3(a)) changes the slope. The threshold value of $P$ is expectedly the highest for Gr/PMMA(0.2) which has the thickest PMMA layer and hence the lowest volume fraction of graphene.

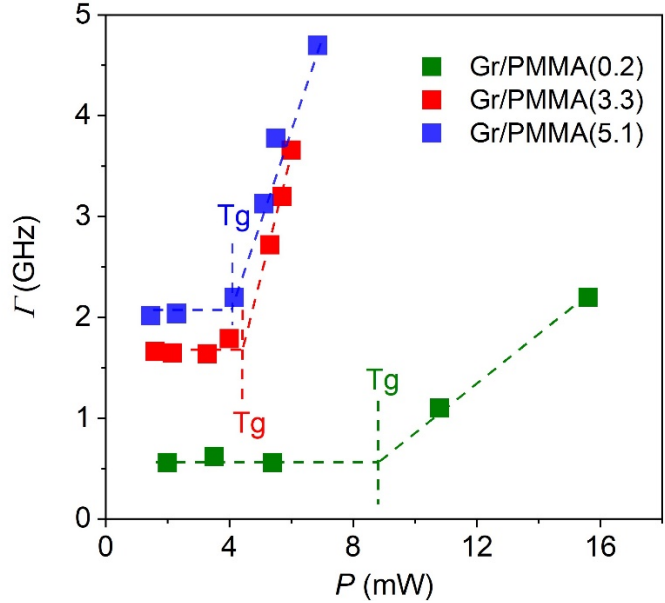

**Figure S9.** The linewidth $\Gamma$ of the BLS peaks recorded in the backscattering geometry as a function of the incident laser power $P$ for the three hybrid Gr/PMMA stacks. The sudden increase of $\Gamma$ at a threshold power occurs when the film temperature reaches the $T_g$ of PMMA, as a result of the laser induced heating beyond a threshold. The dashed lines are guides to the eye, whereas the dashed vertical lines indicate the incident laser power, at which the heating is sufficient to induce glass transition of the PMMA layers.

## 5. Band structure of Gr/PMMA(5.1)

The phononic band structures were calculated by finite element simulations using the COMSOL Multiphysics package. We did eigenfrequency studies by using the Solid Mechanics module. Because of the much larger lateral sizes than the thickness of the sample and the Gr/PMMA periodicity in the thickness direction, we considered a rectangular simulation domain that includes a single layer of graphene and a single layer of PMMA. The thickness of the graphene layer was set to be 0.335 nm, whereas that of the PMMA layer was varied from 3.125 to 100 nm (we did not consider the Gr/PMMA(0.2) sample that has only 6 layers of PMMA). The width of the simulation domain was set to be 1 nm in all simulations. We point out that the value of the width does not affect the calculated in-plane longitudinal and transverse sound velocities in the long wavelength limit, but a small width avoids unnecessary band-folding in the frequency range of interest to this study. We applied periodic boundary conditions on the four sides of the simulation domain. The material properties of graphene and PMMA that are needed for the calculations are listed in figure S10. We discretized the simulation domain by using rectangular elements and verified that the adopted mesh is sufficiently fine so that the obtained eigenfrequencies differ by less than 0.1%, as the number of mesh elements was doubled. We calculated the phonon dispersions in the cross-plane and in-plane directions. The longitudinal and transverse sound velocities were calculated by linear fits to the lowest two branches in the reduced wave vector range of 0-0.005. The wave vectors in the cross-plane and in-plane directions are normalized by $\pi/(d_1 + d_2)$ and $\pi/\text{width}$, respectively. The meanings of $d_1$, $d_2$, and width are shown in figure S10.

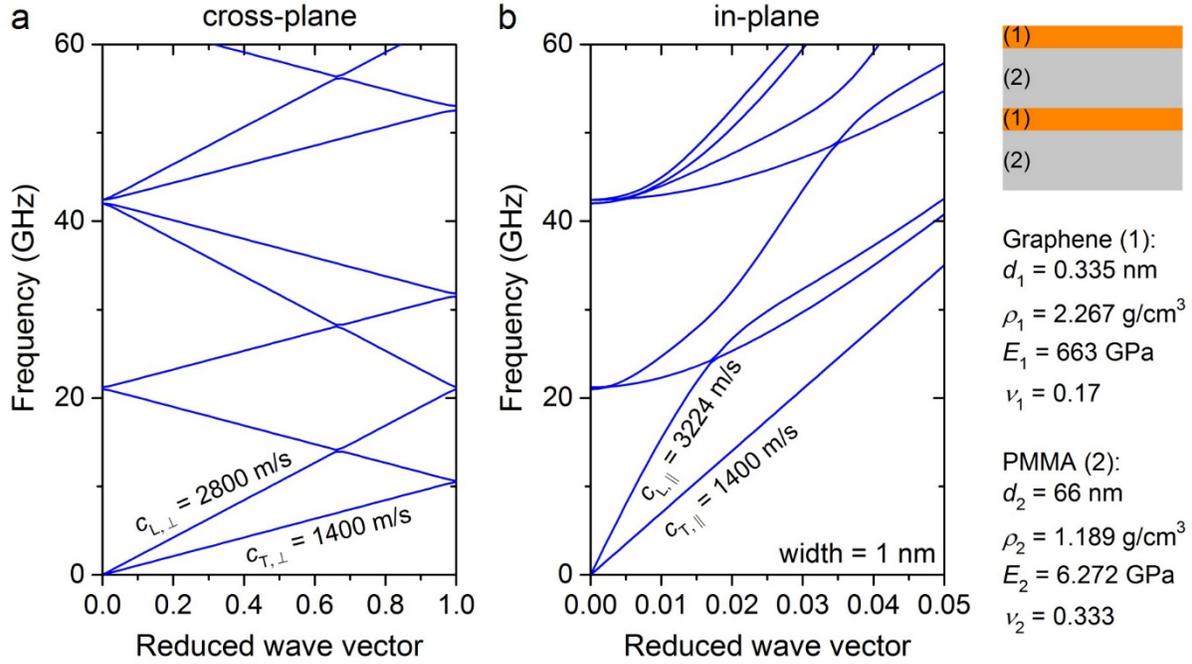

**Figure S10.** Calculated phononic band structure of Gr/PMMA(5.1) in the (a) cross-plane and (b) in-plane directions. The effective sound velocities of the lowest longitudinal and transverse branches in the long wavelength limit are indicated. The schematic on the right illustrates the hybrid Bragg stack. The thicknesses of the graphene and PMMA layers and the material properties are also listed.

We adjusted the Poisson's ratio of graphene, and the corresponding Young's modulus are as follows:

Graphene/PMMA-M: $d_{\text{PMMA}} = 100$ nm

Graphene: $v = 0.07$, $E = 715$ GPa

$v = 0.17$, $E = 696$ GPa

$v = 0.27$, $E = 663$ GPa

Graphene/PMMA-H: $d_{\text{PMMA}} = 66$ nm

Graphene: $v = 0.07$, $E = 681$ GPa

$v = 0.17$, $E = 663$ GPa

$v = 0.27$, $E = 631$ GPa

When $v$ is varied from 0.07 to 0.27, $E$ changes by less than 5%.

## 6. Determination of Graphene's effective modulus in Gr/PMMA Bragg stacks via tensile experiments

We proceed to extrapolation based on the fact that the values in the slopes are in broad agreement with the reported upper bound values of the modulus of elasticity for monolayer CVD graphene (100%vol graphene); as estimated through a wide range of experimental techniques (Table S1). In Figure S10, the modulus of elasticity of Gr/PMMA nanolaminates in graphene contents up to 0.5vol% is presented; we applied the Rule of Mixtures for the Young's modulus as follows:

$$E_c = E_m(1 - \varphi_{Gr}) + \varphi_{Gr} E_{Gr} \Rightarrow$$

$$E_c = E_m - E_m \varphi_{Gr} + \varphi_{Gr} E_{Gr} \Rightarrow$$

$$E_c = E_m + \varphi_{Gr}(E_{Gr} - E_m)$$

Where $E_c$, $E_m$, $E_{Gr}$, $\varphi_{Gr}$ are the composite modulus, the matrix modulus, the graphene modulus and the volume fraction of graphene respectively. The above equation is of the form $y = a + bx$ and therefore in the data presented in Figure S8 we applied an LSQ fit. The matrix modulus at the 100%vol graphene must be included because in the rule of mixtures the slope represents: $E_{gr} - E_{PMMA} = b$, and the modulus of graphene is measured as $E_{gr} = E_{PMMA} + b$.

Based on the above, the effective contribution of graphene's modulus elasticity at the Gr/PMMA Bragg stacks through the mechanical measurements is estimated at $817 \pm 23$ GPa.

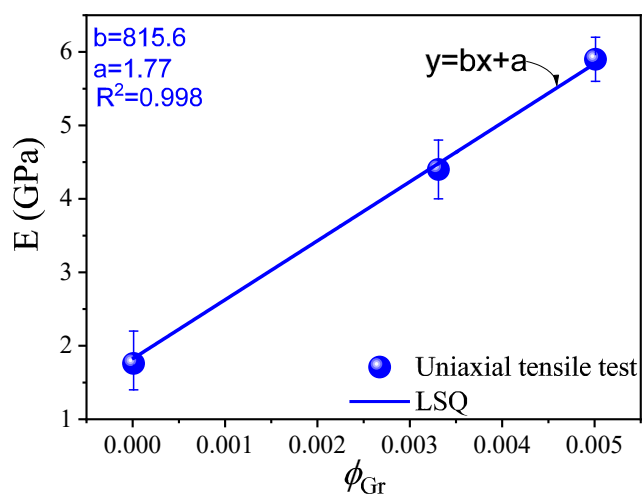

**Figure S11.** Modulus of elasticity in Gr/PMMA Bragg stacks as obtained by uniaxial tensile testing

**Table S1.** Young modulus of CVD graphene as estimated by various experimental techniques.

| Ref. | Method | Young Modulus (GPa) | Graphene condition (Freestanding (F) or on substrate (S)) | Comments |
|---|---|---|---|---|
| [3] | tensile | 950±50 | F | Uniaxial tensile test on single layer CVD Graphene |
| [4] | Nanoindentation | 1000 ± 42.9 | F | Cornerstone paper |
| [5] | Tensile | 637-793 | F | 100 layer polycrystalline CVD graphene |
| | | 728-908 | F | 100 layer near single crystalline CVD graphene Parallel direction |
| | | 683-775 | F | 100 layer near single crystalline CVD graphene Perpendicular direction |
| | | 737 | S | Effective modulus of Model PC/Gr nanocomposite |
| [6] | Bubble inflation | 550 | S | PEMA/Graphene nano-sandwich |
| [7] | Uniaxial tensile test | 1200 ± 500 | S | Low volume fraction Gr/PMMA 0.13%vol nanolaminates |
| [8] | 3-point-bend | 800 | S | Gr/PC folded laminate |
| [9] | Micro-blister Inflation | 751 | S | Nanocomposite Parylene graphene |
| [10] | Nanoindentation | 181 | F | Elasticity softening due to ripples |
| [11] | Nanoindentation | 423 | F | |
| [12] | bulge | 380-950 | F | Grain boundary investigation on the mechanical properties |
| [13] | SEM &XRD | 800 | S | Tube radius examined with SEM and XRD and fitted with continuum elasticity theory |
| [14] | Phonon dispersion (HREELS) | 939-1032 | S | High-resolution electron energy loss spectroscopy (HREELS) experiments were performed using an electron energy loss spectrometer ( |
| [15] | Nanoindentation | 250-1000 | F | Wrinkles softening examination |
| [16] | Interferometric profilometry | Up to 909 | S | The effect of intrinsic crumpling on the mechanics of free-standing graphene |
| [17] | Nanoindentation | 250 | F | |
| [18] | Nanoindentation | 1120 | F | |
| [19] | Nanoindentation | 1000-1500 | F | Depending on the grain size and density |
| [20] | Nanoindentation | 940 | F | Ultra-flat CVD graphene |

**Reference**

[1] S. Mattana, S. Caponi, F. Tamagnini, D. Fioretto, F. Palombo, Viscoelasticity of amyloid plaques in transgenic mouse brain studied by Brillouin microspectroscopy and correlative Raman analysis, J Innov Opt Health Sci 10(6) (2017).
[2] M. Mattarelli, M. Vassalli, S. Caponi, Relevant Length Scales in Brillouin Imaging of Biomaterials: The Interplay between Phonons Propagation and Light Focalization, ACS Photonics 7(9) (2020) 2319-2328.
[3] K. Cao, S. Feng, Y. Han, L. Gao, T. Hue Ly, Z. Xu, Y. Lu, Elastic straining of free-standing monolayer graphene, Nature Communications 11(1) (2020) 284.
[4] G.-H. Lee, R.C. Cooper, S.J. An, S. Lee, A. van der Zande, N. Petrone, A.G. Hammerberg, C. Lee, B. Crawford, W. Oliver, J.W. Kysar, J. Hone, High-Strength Chemical-Vapor–Deposited Graphene and Grain Boundaries, Science 340(6136) (2013) 1073-1076.
[5] B. Wang, D. Luo, Z. Li, Y. Kwon, M. Wang, M. Goo, S. Jin, M. Huang, Y. Shen, H. Shi, F. Ding, R.S. Ruoff, Camphor-Enabled Transfer and Mechanical Testing of Centimeter-Scale Ultrathin Films, Advanced Materials 30(28) (2018) 1800888.
[6] X. Li, J. Warzywoda, G.B. McKenna, Mechanical responses of a polymer graphene-sheet nano-sandwich, Polymer 55(19) (2014) 4976-4982.
[7] I. Vlassiouk, G. Polizos, R. Cooper, I. Ivanov, J.K. Keum, F. Paulauskas, P. Datskos, S. Smirnov, Strong and Electrically Conductive Graphene-Based Composite Fibers and Laminates, ACS Applied Materials & Interfaces 7(20) (2015) 10702-10709.
[8] B. Wang, Z. Li, C. Wang, S. Signetti, B.V. Cunning, X. Wu, Y. Huang, Y. Jiang, H. Shi, S. Ryu, N.M. Pugno, R.S. Ruoff, Folding Large Graphene-on-Polymer Films Yields Laminated Composites with Enhanced Mechanical Performance, Advanced Materials 30(35) (2018) 1707449.
[9] C.N. Berger, M. Dirschka, A. Vijayaraghavan, Ultra-thin graphene–polymer heterostructure membranes, Nanoscale 8(41) (2016) 17928-17939.
[10] C.S. Ruiz-Vargas, H.L. Zhuang, P.Y. Huang, A.M. van der Zande, S. Garg, P.L. McEuen, D.A. Muller, R.G. Hennig, J. Park, Softened Elastic Response and Unzipping in Chemical Vapor Deposition Graphene Membranes, Nano Letters 11(6) (2011) 2259-2263.
[11] Q.-Y. Lin, Y.-H. Zeng, D. Liu, G.Y. Jing, Z.-M. Liao, D. Yu, Step-by-Step Fracture of Two-Layer Stacked Graphene Membranes, ACS Nano 8(10) (2014) 10246-10251.

[12] J.W. Suk, Y. Hao, K.M. Liechti, R.S. Ruoff, Impact of Grain Boundaries on the Elastic Behavior of Transferred Polycrystalline Graphene, Chemistry of Materials 32(14) (2020) 6078-6084.

[13] I.D. Barcelos, L.A.B. Marçal, C. Deneke, L.G. Moura, R.G. Lacerda, A. Malachias, Direct evaluation of CVD multilayer graphene elastic properties, RSC Advances 6(105) (2016) 103707-103713.

[14] A. Politano, G. Chiarello, Probing the Young's modulus and Poisson's ratio in graphene/metal interfaces and graphite: a comparative study, Nano Research 8(6) (2015) 1847-1856.

[15] Q.-Y. Lin, G. Jing, Y.-B. Zhou, Y.-F. Wang, J. Meng, Y.-Q. Bie, D.-P. Yu, Z.-M. Liao, Stretch-Induced Stiffness Enhancement of Graphene Grown by Chemical Vapor Deposition, ACS Nano 7(2) (2013) 1171-1177.

[16] R.J.T. Nicholl, H.J. Conley, N.V. Lavrik, I. Vlassiouk, Y.S. Puzyrev, V.P. Sreenivas, S.T. Pantelides, K.I. Bolotin, The effect of intrinsic crumpling on the mechanics of free-standing graphene, Nature Communications 6(1) (2015) 8789.

[17] C. Gómez-Navarro, M. Burghard, K. Kern, Elastic Properties of Chemically Derived Single Graphene Sheets, Nano Letters 8(7) (2008) 2045-2049.

[18] M. Annamalai, S. Mathew, M. Jamali, D. Zhan, M. Palaniapan, Elastic and nonlinear response of nanomechanical graphene devices, Journal of Micromechanics and Microengineering 22(10) (2012) 105024.

[19] J. Xu, G. Yuan, Q. Zhu, J. Wang, S. Tang, L. Gao, Enhancing the Strength of Graphene by a Denser Grain Boundary, ACS Nano 12(5) (2018) 4529-4535.

[20] B. Deng, Y. Hou, Y. Liu, T. Khodkov, S. Goossens, J. Tang, Y. Wang, R. Yan, Y. Du, F.H.L. Koppens, X. Wei, Z. Zhang, Z. Liu, H. Peng, Growth of Ultraflat Graphene with Greatly Enhanced Mechanical Properties, Nano Letters 20(9) (2020) 6798-6806.